\documentclass[letter,twocolumn]{jpsj3}
\usepackage{graphicx,color} 

\def\diff{\mathrm d}
\newcommand{\etal}{{\it et al.}\ }

\title{Pauli Equation on a Curved Surface and Rashba Splitting on a Corrugated Surface}
\author{Taichi Kosugi$^{1,2}$}

\inst{
$^1$ Nanosystem Research Institute (NRI) ``RICS'', AIST, Umezono, Tsukuba 305-8568, Japan \\
$^2$ Department of Physics, University of Tokyo, Hongo, Tokyo 113-0033, Japan \\
}

\abst{
The Schr\"odinger equation for a spinless charged particle on a curved surface under an electromagnetic field has been obtained by adopting a proper gauge
which allows the separation of the on-surface and transverse dynamics.
[Phys. Rev. Lett. {\bf 100} (2008) 230403]
As its extension, I provide the Pauli equation for a charged spin-$1/2$ particle confined to a curved surface under an electromagnetic field.
Energy spectra of a sphere and a corrugated surface to which a particle is confined are given as simple applications of the equation.
The energy levels obtained exhibit splittings due to the relativistic effect known as the Rashba effect.
}

\kword {Pauli equation, curved surface, spin-orbit interaction, Rashba splitting}

\begin{document}
\maketitle

Much attention has been paid to low-dimensional systems because they show exotic phenomena~\cite{bib:1645_2,bib:1645_3,bib:1655,bib:1657,bib:1650,bib:1680,bib:1683,bib:1647},
essentially distinct from solid state materials.
Recent development of nanotechnology facilitates construction of nanostructures with curved geometry~\cite{bib:1685,bib:1680_3,bib:1680_4,bib:1680_9,bib:1680_13,bib:1680_15}
and those systems can be experimental platforms for study of the interesting phenomena.
Curved geometry induces the geometric potential, which affects the dynamics of an electron moving on the curved surface,
even when an electrostatic potential is absent.

The Schr\"odinger equations on curved surfaces have been used as tools for theoretical investigations on such systems.
Their formulations have been based on the two methods.
The one was proposed by DeWitt~\cite{bib:1681} and the other by da Costa~\cite{bib:1649}.
While the former method regards a curved surface to be a fully two-dimensional space and starts from a Lagrangian for a curved space,
the latter method regards the curved surface as a two-dimensional system embedded in the flat three-dimensional space.
The method established by da Costa, which is called the thin-layer method, has been widely used despite its impossibility of inclusion of arbitrarily oriented magnetic fields.
Da Costa also provided the one-dimensional Schr\"odinger equation for a curved thin tube.
Along this line, Takagi \etal~\cite{bib:1689} studied the geometry-induced Aharonov-Bohm effect by taking into account the torsion of a tube.

Ferrari \etal~\cite{bib:1645} recently adopted the thin-layer approach and rigorously demonstrated by choosing a proper gauge that the separation of the on-surface and transverse dynamics under an electromagnetic field is possible without approximations.
Theoretical investigations on nanostructures are thus expected to treat more various situations in future.

It should be pointed out here that there also exists another rich field of studies for formulation of quantum mechanics on a more general constrained system by employing the possibility of inequivalent quantizations (e.g. Refs. 19-22).
Ohnuki \etal~\cite{bib:1770} analyzed quantum mechanics on $S^D$ embedded in $(D+1)$-dimensional flat space by setting up appropriate fundamental algebra,
with gauge potentials induced.

Parallel to the extension of the possibility of geometric arrangement,
the role of relativistic effects such as spin-orbit interaction has been increasingly evoking interests in material physics due to the recent rise of spintronics.
The relativistic effects and the spin degree of freedom for an electronic system naturally emerge in the Dirac equation~\cite{bib:Sakurai_Advanced}.
Since the upper two components of a four-component spinor are much larger than the lower two components
in ordinary condensed matter physics,
the large part of relativistic electronic structure calculations have been performed using two-component wave functions.
The most tractable tool for a quantum mechanical analysis on an electronic system including magnetic properties and relativistic effects is hence the Pauli equation,
which is for a charged spin-$1/2$ particle with a nonzero mass and contains the lowest-order relativistic correction term.
It will provide deep insights into rich physics on nanosystems.

In this Letter I provide the Pauli equation for a particle confined to a curved surface under an electromagnetic field
as an extension of the work done by Ferrari \etal~\cite{bib:1645}
I reconcile ourselves to two assumptions mentioned below for performing variable separation in the present study.
Excepting those assumptions, the derivation of the Pauli equation will proceed on the same strictness as for the Schr\"odinger case.

The expansion of the Dirac equation for a spin-$1/2$ particle of its mass $m$ and charge $Q$ under an electromagnetic field using the Foldy-Wouthuysen method~\cite{bib:152} leads to
the Pauli equation $i \frac{\partial \psi}{\partial t} = H_{\mathrm{P}} \psi$ for the upper two components $\psi$ of the four-component spinor.
$H_{\mathrm{P}}$ is the Pauli Hamiltonian,
which neglects the mass term $mc^2$ and the terms on the orders higher than $m^{-2}$, given by
\begin{gather}
	H_{\mathrm{P}} =
	\frac{\boldsymbol{\Pi}^2}{2m} + V
	- \frac{Q}{mc} \boldsymbol{S} \cdot \boldsymbol{B}
	\nonumber \\
	- \frac{Q}{4m^2c^2} [
		\boldsymbol{\Pi} \cdot \boldsymbol{S} \times \boldsymbol{E}
		+ \boldsymbol{S} \times \boldsymbol{E} \cdot \boldsymbol{\Pi}
	]
	.
	\label{Pauli_Cartesian}
\end{gather}
$\boldsymbol{\Pi} \equiv -i \nabla - \frac{Q}{c} \boldsymbol{A}$ is the canonical momentum operator and the magnetic field $\boldsymbol{B} = \nabla \times \boldsymbol{A}$ is the rotation of the vector potential.
$\boldsymbol{S} = \boldsymbol{\sigma}/2$ is the spin operator and $\boldsymbol{\sigma}$ is the Pauli matrix.
The contribution from the divergence of the electric field $\nabla \cdot \boldsymbol{E}$, called the Darwin term, is absorbed into the scalar potential $V$.
All spin-orbit interactions which will appear below come from the last term on the right hand side of eq. (\ref{Pauli_Cartesian}).
Let us consider a coordinate transformation from the Cartesian coordinates $x_{(a)} (a=x,y,z)$ to the curvilinear coordinates $q_i (i=1,2,3)$.
The letter in a parenthesis stands for the Cartesian coordinate and not in a parenthesis for the new coordinate.
The dreibein field and its inverse are defined as
\begin{gather}
	e_i^{(a)} = \frac{\partial x_{(a)}}{\partial q_i} , \
	e_{(a)}^i = \frac{\partial q_i}{\partial x_{(a)}} 
	,
\end{gather}
which satisfy the conditions $e_i^{(a)} e_{(a)}^j = \delta_i^j , e_i^{(a)} e_{(b)}^i = \delta_{(b)}^{(a)}$ due to the chain rule of derivative.
Summation is implied over the repeated index.
The metric tensor in the new coordinate system is given by $G_{ij} = e_i^{(a)} e_j^{(a)}$.
We define $v_i \equiv e_i^{(a)} v_{(a)}$ for a vector $\boldsymbol{v} = v_{(a)} \boldsymbol{e}_{(a)}$, where $\boldsymbol{e}_{(a)}$ is the unit vector along $a$ direction in the Cartesian coordinate system,
from which it follows that $v_{(a)}=e_{(a)}^i v_i$.
Using the Laplacian for the curvilinear coordinate system~\cite{bib:Arfken} and the covariant derivative $D_i \equiv \partial_i - \frac{iQ}{c} A_i$,
the Pauli Hamiltonian, eq. (\ref{Pauli_Cartesian}), is rewritten as
\begin{gather}
	H_{\mathrm{P}} = - \frac{1}{2m} \frac{1}{\sqrt{G}} D_i \sqrt{G} G^{ij} D_j + V
		- \frac{Q}{2mc} \sigma_{(a)} \widetilde{B}_{(a)}
	\nonumber \\
		+ \frac{iQ}{4m^2c^2} h^{ij} E_i D_j
	,
\end{gather}
where $G \equiv \det G_{ij}$ and $G^{ij}$ is the inverse metric tensor.
We have defined $\widetilde{\boldsymbol{B}} \equiv \boldsymbol{B} - \frac{i}{4mc} \frac{\partial \boldsymbol{B}}{\partial t}$
using the Maxwell's equation $\nabla \times \boldsymbol{E} = - \frac{1}{c} \frac{\partial \boldsymbol{B}}{\partial t}$.
We have defined the matrix $h^{ij} \equiv \varepsilon_{(abc)} \sigma_{(a)} e_{(b)}^i e_{(c)}^j = -h^{ji}$, where $\varepsilon_{(abc)}$ is the Levi-Civita symbol.
The Pauli equation is obviously invariant under the following gauge transformation with an arbitrary scalar function $\gamma$~\cite{bib:1645}:
\begin{gather}
	V \to V' = V - \frac{Q}{c} \frac{\partial \gamma}{\partial t}, \nonumber \\
	A_i \to A_i' = A_i + \partial_i \gamma, \nonumber \\
	\psi \to \psi' = e^{iQ \gamma/  c} \psi
	.
\end{gather}
Writing explicitly the spatial derivative with respect to the new coordinates, we write down the Pauli equation as
\begin{gather}
	i \frac{\partial \psi}{\partial t} = - \frac{1}{2m}
		\Bigg[
			\frac{1}{\sqrt{G}} \partial_i (\sqrt{G} G^{ij} \partial_j \psi) - \frac{iQ}{ c} \frac{1}{\sqrt{G}} \partial_i ( \sqrt{G} G^{ij} A_j) \psi
			\nonumber \\
			- \frac{2iQ}{ c} G^{ij} A_i \partial_j \psi - \frac{Q^2}{ c^2} G^{ij} A_i A_j \psi
		\Bigg]
	\nonumber \\
		+ V \psi
		- \frac{Q}{2mc} \sigma_{(a)} \widetilde{B}_{(a)} \psi
		+ \frac{iQ}{4m^2c^2} h^{ij} E_i D_j \psi
	.
	\label{Eq_psi}
\end{gather}
From here the curved surface $S$ confining the particle is considered.
Let us adopt a coordinate transformation such that 
$S$ is described as $q_3 = 0$
and an arbitrary point $\boldsymbol{r}$ 
immediately close to the point $\boldsymbol{r}_S$ on $S$ is given by
$\boldsymbol{r}(q_1, q_2, q_3) = \boldsymbol{r}_S(q_1, q_2) + q_3 \boldsymbol{n}(q_1, q_2)$,
where $\boldsymbol{e}_a \equiv \frac{\partial \boldsymbol{r}_S}{\partial q_a} (a=1,2)$ and
$\boldsymbol{n} \equiv \boldsymbol{e}_1 \times \boldsymbol{e}_2 / |\boldsymbol{e}_1 \times \boldsymbol{e}_2|$ is the unit normal vector of $S$ at $\boldsymbol{r}_S$.
It obviously follows that
$e^{(a)}_i = \boldsymbol{e}_i \cdot \boldsymbol{e}_{(a)} (i=1,2), 
e^{(a)}_3 = \boldsymbol{n} \cdot \boldsymbol{e}_{(a)}$ on $S$.
The two-dimensional induced metric tensor $g_{ab} = \boldsymbol{e}_a \cdot \boldsymbol{e}_b$
is connected with the three-dimensional one via the following relation:
\begin{gather}
	G_{ab} = g_{ab} + [ \alpha g + {}^\mathrm{t}(\alpha g)]_{ab} q_3 + (\alpha g {}^\mathrm{t} \alpha)_{ab} q_3^2, \nonumber \\
	G_{a3} = G_{3a} = 0, \ G_{33} = 1
	.
\end{gather}
$\alpha_{ab}$ is the Weingarten matrix~\cite{bib:1649}, which satisfies
$\frac{\partial \boldsymbol{n}}{\partial q_a} = \alpha_{ab} \boldsymbol{e}_b$.
As is done in the case of the Schr\"odinger equation~\cite{bib:1649,bib:1645,bib:1650},
we put the wave function in the form
\begin{gather}
	\psi(q_1, q_2, q_3, t) = 
		\frac{\chi(q_1, q_2, q_3, t)}{\sqrt{1 + \mathrm{Tr} \, \alpha q_3 + \det \alpha q_3^2 }} 
	,
	\label{def_chi}
\end{gather}
which ensures the norm conservation condition 
$\int |\psi|^2 \sqrt{G} \diff^3q = \int |\chi|^2 \sqrt{g} \diff^3q$.
Substituting eq. (\ref{def_chi}) into eq. (\ref{Eq_psi}) and
taking the limit $q_3 \to 0$, we obtain the equation for $\chi$:
\begin{gather}
	i \frac{\partial \chi}{\partial t} = - \frac{1}{2m}
		\Bigg[
			\frac{1}{\sqrt{g}} \partial_a (\sqrt{g} g^{ab} \partial_b \chi) - \frac{iQ}{ c} \frac{1}{\sqrt{g}} \partial_a ( \sqrt{g} g^{ab} A_b) \chi
			\nonumber \\
			- \frac{2iQ}{ c} g^{ab} A_a \partial_b \chi - \frac{Q^2}{ c^2} (g^{ab} A_a A_b + A_3^2 ) \chi
			\nonumber \\
			+ \partial_3^2 \chi - \frac{iQ}{ c} (\partial_3 A_3) \chi - \frac{2iQ}{ c} A_3 \partial_3 \chi
		\Bigg]
		+ V_S \chi 
		\nonumber \\
		+ V \chi
		- \frac{Q}{2mc} \sigma_{(a)} \widetilde{B}_{(a)} \chi
	\nonumber \\
		+ \frac{iQ}{4m^2c^2} \Bigg( h^{ij} E_i D_j - \frac{1}{2} \mathrm{Tr} \, \alpha  h^{i3} E_i \Bigg) \chi
	,
	\label{Eq_chi1}
\end{gather}
where $V_S(q_1, q_2) = -\frac{1}{2m} [ (\mathrm{Tr} \, \alpha)^2/4 - \det \alpha ]$
is the well-known geometric potential.
The electromagnetic field is evaluated at $q_3 = 0$.
We here apply a gauge transformation with a scalar function
$\gamma(q_1, q_2, q_3) \equiv - \int_0^{q_3} A_3(q_1, q_2, z) \diff z$,
as done in the Schr\"odinger case~\cite{bib:1645},
so that $A_3'$ and $\partial_3 A_3'$ vanish and $A_1$ and $A_2$ are unchanged on $S$.
We divide the scalar potential into the on-surface electric part and the confinement part as
$V = V_{\mathrm{el}}(q_1, q_2) + V_{\lambda}(q_3)$.
The gauge-transformed equation reads, from eq. (\ref{Eq_chi1}),
\begin{gather}
	i \frac{\partial \chi}{\partial t} = - \frac{1}{2m}
		\Bigg[
			\frac{1}{\sqrt{g}} \partial_a (\sqrt{g} g^{ab} \partial_b \chi) - \frac{iQ}{ c} \frac{1}{\sqrt{g}} \partial_a ( \sqrt{g} g^{ab} A_b) \chi
			\nonumber \\
			- \frac{2iQ}{ c} g^{ab} A_a \partial_b \chi - \frac{Q^2}{ c^2} g^{ab} A_a A_b  \chi
		\Bigg]
		+ V_S \chi 
		\nonumber \\
		+ V_{\mathrm{el}} \chi
		- \frac{Q}{2mc} \sigma_{(a)}  \widetilde{B}_{(a)} \chi
		+ \frac{iQ}{4m^2c^2} h^{ia} E_i D_a  \chi
		+ H_3 \chi
	,
	\label{diff_eq_chi}
\end{gather}
where
\begin{gather}
	H_3 \equiv
		- \frac{1}{2m}  \partial_3^2  + \frac{iQ}{4m^2c^2} h^{i3} E_i ( \partial_3 - \mathrm{Tr} \, \alpha/2 )
		+ V_\lambda 
	\label{Hamiltonian_q3}	
\end{gather}
is the Hamiltonian describing the dynamics along $q_3$ direction.
$H_3$ contains $q_1, q_2$ and $t$ as parameters.
The term involving the first-order derivative on the right hand side couples the on-surface and transverse dynamics as a relativistic effect.
In the case of the Schr\"odinger equation, this term were absent and hence the variable separation would have been done.
Let the solution of an eigenvalue problem $H_3 \phi = \varepsilon \phi$ of the form
\begin{gather}
	\phi(q_1,q_2,q_3, t) = f(q_1, q_2, q_3, t)
	\begin{pmatrix}
		g_\uparrow(q_1, q_2, t) \\
		g_\downarrow(q_1, q_2, t) \\
	\end{pmatrix}
	.
\end{gather}
The condition for a nontrivial solution is then, from eq. (\ref{Hamiltonian_q3}),
\begin{gather}
	\Bigg[ \Bigg(- \frac{\partial_3^2}{2m}  + V_\lambda - \varepsilon \Bigg) f \Bigg]^2
	+ \Bigg[ \frac{Q \tilde{E}}{4 m^2 c^2}  \Bigg( \partial_3 - \frac{\mathrm{Tr} \, \alpha}{2} \Bigg) f \Bigg]^2 = 0
	\label{nontrivial}
\end{gather}
with $g_\uparrow$ and $g_\downarrow$ arbitrary,
where $\tilde{E}_{(a)} \equiv \varepsilon_{(abc)} e_{(b)}^i e_{(c)}^3 E_i$ and
$\tilde{E} \equiv \sqrt{\tilde{E}_{(x)}^2 + \tilde{E}_{(y)}^2 + \tilde{E}_{(z)}^2 }$.
We try the form $f(q_1, q_2, q_3, t) = \exp (\pm \frac{iQ}{4 mc^2} \tilde{E} q_3 ) \tilde{f}(q_3)$
and substitute it into eq. (\ref{nontrivial}).
The condition then becomes
\begin{gather}
	\Bigg[ - \frac{1}{2m} \partial_3^2 
		+ V_\lambda - \frac{ Q^2 \tilde{E}^2 }{32 m^3 c^4} 
		\mp \frac{i Q \tilde{E}}{8m^2c^2}  \mathrm{Tr} \, \alpha 
	\Bigg] 
	\tilde{f}
	= \varepsilon \tilde{f}
	\label{diff_eq_f}
	.
\end{gather}
It implies that the eigenvalue is expressed as
\begin{gather}
	\varepsilon = \tilde{\varepsilon} - \frac{ Q^2  \tilde{E}^2}{32 m^3 c^4}
	\mp \frac{i Q \tilde{E}}{8m^2c^2}  \mathrm{Tr} \, \alpha 
	,
	\label{evalue_H3}
\end{gather}
where $\tilde{\varepsilon}$ is an eigenvalue for $\tilde{E} = 0$, independent of $q_1, q_2$ and $t$.
We neglect the second term on the right hand side of eq. (\ref{evalue_H3}) since it is on the order of $m^{-3}$.
In addition, we take the average of the positive and negative signatures of the third term so that it vanishes.
The first assumption needed for the present derivation is the validity of this treatment.
The confinement potential is so strong that the wave function at any point on $S$ is, as the second assumption, expected to be written as
\begin{gather}
	\chi(q_1, q_2, q_3, t) = \exp \Bigg(\pm \frac{iQ \tilde{E}}{4 mc^2}  q_3 \Bigg) \tilde{f}_0(q_3) \chi_S(q_1, q_2, t)
	,
	\label{chi_f_chis}
\end{gather}
where $\tilde{f}_0$ is the nondegenerate solution of eq. (\ref{diff_eq_f}) with the lowest eigenvalue $\tilde{\varepsilon}_0$.
This assumption is feasible as long as low-energy excitation is discussed.
The two assumptions introduced above are not needed for the Schr\"odinger case~\cite{bib:1645} thanks to the absence of the relativistic term,
for which the differential equation of the transverse dynamics remains unsolved after the variable separation.
Hereafter we set $\tilde{\varepsilon}_0 \equiv 0$.
Substitution of eq. (\ref{chi_f_chis}) into eq. (\ref{diff_eq_chi}) leads to the equation to be satisfied by the surface wave function $\chi_S$:
\begin{gather}
	i \frac{\partial \chi_S}{\partial t} = - \frac{1}{2m}
		\Bigg[
			\frac{1}{\sqrt{g}} \partial_a (\sqrt{g} g^{ab} \partial_b \chi_S) 
			\nonumber \\
			- \frac{iQ}{ c} \frac{1}{\sqrt{g}} \partial_a ( \sqrt{g} g^{ab} A_b) \chi_S
			- \frac{2iQ}{ c} g^{ab} A_a \partial_b \chi_S
			\nonumber \\
			- \frac{Q^2}{ c^2} g^{ab} A_a A_b  \chi_S
		\Bigg]
		+ ( V_S + V_{\mathrm{el}} ) \chi_S
		+ H_{\mathrm{sp-rel}} \chi_S
	\label{Pauli_eq_chi}
	,
\end{gather}
where
\begin{gather}
	H_{\mathrm{sp-rel}} \equiv
		- \frac{Q}{2mc} \sigma_{(a)}  \widetilde{B}_{(a)}
		+ \frac{iQ}{4m^2c^2} h^{ia} E_i D_a 
	.
\end{gather}
It has been demonstrated that the variable separation is possible with the two assumptions for the Pauli equation on a curved surface under an electromagnetic field.
The Hamiltonian of the resultant equation, eq. (\ref{Pauli_eq_chi}), is simply that for the Schr\"odinger equation on a curved surface obtained by Ferrari \etal~\cite{bib:1645}
plus $H_{\mathrm{sp-rel}}$.
The new term acts on the two components of the surface wave function differently in general and hence physics which is absent in the Schr\"odinger case can emerge.

In what follows, energy spectra of a charged spin-$1/2$ particle on a sphere and a corrugated surface are provided as instructive applications of the surface Pauli equation.
Analyses of their spinless and nonrelativistic electronic properties have been done using the Schr\"odinger equaiton for curved surfaces.\cite{bib:1645_2,bib:1645_3,bib:1680}

As the first example,
let us calculate the energy spectrum of a charged spin-$1/2$ particle confined to a sphere of radius $R$.
We here adopt the spherical coordinates as a new coordinate system.
We assume that there is no magnetic field and a static electric field with a constant amplitude penetrates perpendicularly to the sphere:
$\boldsymbol{E} = E \boldsymbol{e}_r$.
In this case $V_S = V_{\mathrm{el}} = 0, E_r=E, E_\theta = E_\phi = 0$ and 
\begin{gather}
	h^{r \theta} = - \frac{\sin \phi}{R} \sigma_{(x)} + \frac{\cos \phi}{R} \sigma_{(y)} \nonumber \\
	h^{r \phi} =
		-\frac{\cot \theta \cos \phi}{R} \sigma_{(x)}
 		-\frac{\cot \theta \sin \phi}{R} \sigma_{(y)}
		+\frac{1}{R} \sigma_{(z)} \nonumber \\
	h^{\theta \phi} =
		\frac{\cos \phi}{R^2} \sigma_{(x)}
 		+\frac{\sin \phi}{R^2} \sigma_{(y)}
		+\frac{\cot \theta}{R^2} \sigma_{(z)}	
\end{gather}
on the sphere.
The Pauli Hamiltonian on the sphere is thus
\begin{gather}
	H = - \frac{1}{2m R^2}
		\Bigg[ \frac{1}{\sin \theta} \partial_\theta \sin \theta \partial_\theta
			+ \frac{1}{\sin^2 \theta} \partial_\phi^2 
		\Bigg]
	\nonumber \\
		- \frac{i \alpha}{R}
		\begin{pmatrix}
			\partial_\phi & -i e^{-i \phi} ( \partial_\theta - i \cot \theta \partial_\phi) \\
			i e^{i \phi} ( \partial_\theta + i \cot \theta \partial_\phi) & - \partial_\phi \\
		\end{pmatrix}
	\nonumber \\
	= \frac{L^2}{2m R^2} 
		+ \frac{2 \alpha}{R} \boldsymbol{L} \cdot \boldsymbol{S}
	,
	\label{H_sphere}
\end{gather}
where $\alpha \equiv -\frac{QE}{4m^2c^2}$ and $\boldsymbol{L}$ is the orbital angular momentum operator.
This Hamiltonian contains the spin-orbit interaction in the form well known in condensed matter physics,
which is absent in the Schr\"odinger case.
Spinor spherical harmonics~\cite{bib:1042}
\begin{gather}
	\mathcal{Y}_{jl}^{j_z} =
		\begin{pmatrix}
 				\sqrt{\frac{l \pm j_z + 1/2}{2l+1}} Y_{l j_z-1/2} \\
 			\pm \sqrt{\frac{l \mp j_z + 1/2}{2l+1}} Y_{l j_z+1/2}
	\end{pmatrix}
	,
\end{gather}
is a simultaneous eigenfunction of $J^2, J_z, L^2$ and $S^2$ with the eigenvalues
$j(j+1), j_z, l(l+1)$ and $3/4$, respectively.
$\boldsymbol{J} \equiv \boldsymbol{L} + \boldsymbol{S}$ is the total angular momentum operator and $j$ can take only $l \pm 1/2$.
It is easily confirmed that $\mathcal{Y}_{jl}^{j_z}/R$ is a normalized eigenfunction of the Hamiltonian, eq. (\ref{H_sphere}), with the eigenvalue
\begin{gather}
	\varepsilon_{jl} = \frac{l(l+1)}{2m R^2} 
		+ \frac{\alpha}{R} [j(j+1) - l(l+1) - 3/4]
	.
	\label{evalue_sphere}
\end{gather}
The energy gap between two levels with common $l$ and different $j$ is $\frac{\alpha}{R}(2l + 1)$,
which is a relativistic effect.
Since $j \pm 1/2 = l \approx mvR = kR$, where $v$ is the velocity of the particle and $k$ is its wave number,
in the limit of $R \to \infty$ the energy eigenvalue converges to $\varepsilon_{jl} \approx \frac{k^2}{2m} \pm \alpha k$.
This energy dispersion is the same as that for the Rashba Hamiltonian~\cite{bib:Rashba}
$H_{\mathrm{R}} = \frac{k^2}{2m} + \alpha \boldsymbol{\sigma} \cdot \boldsymbol{e}_z \times \boldsymbol{k}$,
which describes a charged particle moving freely on $xy$ plane under an electric field along $z$ axis,
leading to the spin splitting as a relativistic effect due to the lack of inversion symmetry.

As the second example,
let us calculate the energy spectrum of a charged spin-$1/2$ particle confined to a currugated surface.
We assume here that the corrugation is along $x$ direction represented by $z = f(x)$ and there is no magnetic field and a static electric field with a constant amplitude penetrates perpendicularly to the surface:
$\boldsymbol{E} = E (-f' \boldsymbol{e}_x + \boldsymbol{e}_z)/\sqrt{1 + f'^2}$ and $V_{\mathrm{el}} = 0$.
We define the new coordinates on $S$ as
$q_1 = \int_0^x \diff x' \sqrt{1 + f'(x')^2}, q_2 = y, q_3 = 0$.
$q_1$ is the line length along $S$.
$\boldsymbol{e}_1 = ( \boldsymbol{e}_x + f' \boldsymbol{e}_z )/\sqrt{ 1 + f'^2 }, \boldsymbol{e}_2 = \boldsymbol{e}_y$
and thus
$g_{ab} = \delta_{ab}, \boldsymbol{n}	= ( - f' \boldsymbol{e}_x + \boldsymbol{e}_z)/\sqrt{ 1 + f'^2 }$ and
$V_S = - \frac{1}{8m} \frac{f''^2}{(1 + f'^2)^3}$.
The nonzero components of the dreibein on $S$ are
$e^1_{(x)} = e^3_{(z)} = \frac{1}{\sqrt{1 + f'^2}}, e^2_{(y)} = 1, e^1_{(z)} = -e^3_{(x)} = \frac{f'}{\sqrt{1 + f'^2}}$.
Using $E_1 = E_2 = 0, E_3 = E, V_{\mathrm{el}} = 0$ and
\begin{gather}
	h^{12} = \frac{\sigma_{(z)} - f' \sigma_{(x)}}{\sqrt{1 + f'^2}} \nonumber  \\
	h^{23} = \frac{\sigma_{(x)} + f' \sigma_{(z)}}{\sqrt{1 + f'^2}} \nonumber \\
	h^{31} = \sigma_{(y)} 
\end{gather}
on $S$,
we obtain the following Pauli Hamiltonian for the surface wave function:
\begin{gather}
	H =	- \frac{1}{2m} ( \partial_1^2 + \partial_2^2 )
		+ V_S
		\nonumber \\
		- i \alpha
		\Bigg(
			\sigma_{(y)} \partial_1
			- \frac{\sigma_{(x)} + f' \sigma_{(z)}}{\sqrt{1 + f'^2}} \partial_2
		\Bigg)
		,
	\label{H_corrugated}
\end{gather}
where the term proportional to $\alpha$ is responsible for relativistic effects.
We examine periodic corrugation on $S$ in what follows,
putting $f(x) = h g(x)$, where the constant $h$ measures the height of the corrugation and $g$ is a periodic function.
When the corrugation is absent $(h = 0)$, the system reduces to the ordinary Rashba system~\cite{bib:Rashba}
and the time-independent Pauli equation $H u = \varepsilon u$ has two plane wave solutions:
\begin{gather}
	\varepsilon^0_\pm = \frac{k^2}{2m} \pm \alpha k, \, 
	u^0_\pm = \frac{e^{i(k_1 q_1 + k_2 q_2)}}{\sqrt{2}}
	\begin{pmatrix}
		1 \\
		\pm i e^{i \theta_k}
	\end{pmatrix}
	,
\end{gather}
where $k_1$ and $k_2$ are the wave numbers, $k \equiv \sqrt{k_1^2 + k_2^2}$ and $\theta_k$ is the argument of $k_1 + ik_2$ on the complex plane.
From the perturbation theory, the perturbed energy spectrum to the lowest order for small corrugation
is evaluated as the expectation value of the perturbed Hamiltonian with respect to the unperturbed wave functions.
Expansion of the Pauli Hamiltonian, eq. (\ref{H_corrugated}), in terms of $h$
leads to the lowest-order perturbation
\begin{gather}
	H^{(1)} = i \alpha h g' \sigma_{(z)} \partial_2
\end{gather}
and the second lowest-order perturbation
\begin{gather}
	H^{(2)} = - \frac{h^2 g''^2}{8m} - i \frac{ \alpha h^2 g'^2}{2} \sigma_{(x)} \partial_2
	.
\end{gather}
Since $\langle u^0_\pm | H^{(1)} | u^0_\pm \rangle = 0$ due to the periodicity of $g$ and
$\langle u^0_\pm | H^{(2)} | u^0_\pm \rangle = -\frac{h^2 \langle  g''^2 \rangle}{8m} \mp \frac{\alpha }{2} h^2 \langle g'^2 \rangle k \sin^2 \theta_k$,
the perturbed energy dispersion is given by
\begin{gather}
	\varepsilon_\pm = \frac{k^2}{2m} 
		\pm \alpha k \Bigg( 1 - \frac{h^2 \langle g'^2 \rangle}{2} \sin^2 \theta_k \Bigg)
		-\frac{h^2 \langle g''^2 \rangle}{8m}
	,
\end{gather}
which indicates that any small periodic corrugation decreases the Rashba splitting.

In conclusion, I have provided the Pauli equation for a charged spin-$1/2$ particle confined to a curved surface under an electromagnetic field
by performing the variable separation on the manner similar to that for the Schr\"odinger equation.
Energy spectra of a sphere and a corrugated surface to which a particle is confined were calculated.
It was demonstrated that any small periodic corrugation decreases the Rashba splitting on a surface.
The basic equation obtained will be a powerful tool for analyses of physics on curved surfaces and nanostructures
where spin degree of freedom and/or spin-orbit interactions play important roles.

\begin{acknowledgement}
The author would like to thank Takashi Miyake and Shoji Ishibashi for useful discussions.
This work was partly supported by the Next Generation Super Computing
Project, Nanoscience Program,
and by a Grant-in-Aid for Scientific Research on Innovative Areas,
"Materials Design through Computics: Complex Correlation and Non-Equilibrium Dynamics" (No. 22104010) from MEXT, Japan.
\end{acknowledgement}

\end{document}